\documentclass[preprint,12pt]{article}
\usepackage{graphicx}
\usepackage{color}
\usepackage{amssymb,amsthm}
\usepackage{float}
\usepackage{perpage} 
\MakePerPage{footnote} 

\begin{document}
{\large \bf Dragon-kings death in nonlinear wave interactions}\\

\noindent Moises S. Santos$^{1}$, Jos\'e D. Szezech Jr$^{1,2}$, Antonio M.
  Batista$^{1,2}$, Kelly C. Iarosz$^3$, Iber\^e L. Caldas$^3$, Ricardo L. Viana$^4$\\

\noindent $^1$Graduate in Science Program - Physics, State University of Ponta Grossa, 84030-900, Ponta Grossa, PR, Brazil.\\
$^2$Department of Mathematics and Statistics, State University of Ponta Grossa, 84030-900, Ponta Grossa, PR, Brazil.\\
$^3$Institute of Physics, University of S\~ao Paulo, 05508-900, S\~ao Paulo, SP, Brazil.\\
$^4$Department of Physics, Federal University of Paran\'a, 80060-000, Curitiba, PR, Brazil.\\
\noindent Corresponding author: moises.wz@gmail.com\\

\noindent {\bf Abstract} \\ 
\noindent Extreme events are characterised by low probabilities and high impact on the
systems. As a consequence, it is possible to find many studies about the
predictability and suppression of extreme events. In this work, we show the
existence of dragon-kings extreme events in nonlinear three-wave interactions.
Dragon-king extreme events, identified by phase transitions, tipping points and
catastrophes, affects fluctuating systems. We show that these events can be
avoided by adding a perturbing small amplitude wave to the system.

\section{Introduction}
Extreme events are by definition scarce and they can have a relevant impact on
the systems. These events have been observed in weather \cite{sura11}, optics
\cite{clerc16}, and plasma physics \cite{piet14}. In optical rogue waves,
extreme events emerge from a turbulent state \cite{akhmediev16}. Riccardo et al.
\cite{riccardo04} reported that extreme event is compatible with general halo
current trends in magnetically confined plasma physics in the tokamak Joint
European Torus (JET).
 
Extreme events have been estimated through the extrapolation of power law
frequency-size distributions \cite{sachs12}. Many extreme events, known as
dragon-kings, do not belong to a power law distribution \cite{sornette09}.
The dragon-kings were reported by Johansen and Sornette \cite{johansen1998} in
the distribution of financial drawdowns. Sornette and Ouillon \cite{sornette12}
discussed the mechanisms, statistical tests, and empirical evidences of
dragon-kings. Dragon-king extreme events was found in neuronal networks by
Mishra et al. \cite{mishra18}. They presented evidence of these events in
coupled bursting neurons.

Recently, many articles about the suppression of extreme events have been
published. Direct corrective reset to the system was used as a suppression
method \cite{munt14}. Cavalcante et al. \cite{cavalcante13} studied coupled
chaotic oscillators and showed that extreme events can be suppressed by means
of tiny perturbations. Small perturbations were used to reduce the appearance
of extreme events in damped one-dimensional nonlinear Schr\"odinger equation
\cite{galuzio14}. Kr\"uger et al. \cite{kruger15} demonstrated that the
existence of extreme events can be controlled by means of the properties of
the phase space in Hamiltonian systems.

We analyse a nonlinear three-wave interactions model. The nonlinear three-wave
interaction is the lowest order effect in systems described by waves
superposition \cite{yaakobi08}. It plays an important role in nonlinear optics,
plasma physics and hydrodynamics \cite{kaup79}. Chian and Abalde \cite{chian99}
developed a nonlinear theory of three-wave interactions of Langmuir waves with
whistler waves in the solar wind. Batista et al. \cite{batista06} considered
three-wave interaction to investigate drift-wave turbulence in tokamak edge
plasma. In our simulations, we identify Lorentzian pulses. These pulses have
been observed in magnetised plasmas and interactions of drift-Alfv\'en waves
\cite{pace08}. They can appear in flow trajectories due to topological
alterations and in chaotic orbits \cite{maggs12a,maggs12b}.

We calculate the frequency-size distributions of the wa\-ve amplitude in the
three-wave interactions model. Depending on the parameter, the distribution
follows a power law except for large sizes. This behaviour indicates the
existence of dragon-kings extreme events. Aiming to suppress the
dragon-kings, we propose a fourth wave as a control method. Batista et al.
\cite{batista08} used a fourth resonant wave of small amplitude to control
chaotic behaviour of three interacting modes. Coexistence of attractors was
observed in a dissipative nonlinear parametric four-wave interactions
\cite{coninck04}. In this work, we show that a fourth wave is able to kill the
dragon-kings in nonlinear three-wave interactions. 

The paper is organised as follows: In Section $2$, we describe the nonlinear
three-wave interactions model. Section $3$ shows the dragon-kings extreme
events and presents our method to suppress these events. In the last Section,
we draw our conclusions.

\section{Nonlinear tree-wave interactions model}
The nonlinear three-wave interactions are described by three first-order
differential equations \cite{wersing80,meunier82,lopes96}. We include a fourth
wave interacting with the first and second waves \cite{coninck04}. The equations
of the four-wave interactions are given by
\begin{eqnarray}\label{eq1}
  \dot{A}_1 & = & \nu_1A_1+A_2A_3-rA_2^*A_4,\\
  \dot{A}_2 & = & \nu_2A_2-A_1A_3^*-rA_1^*A_4,\\
  \dot{A}_3 & = & \nu_3A_3-A_1A_2^*+{\rm i}\delta_3A_3,\\
  \dot{A}_4 & = & \nu_4A_4-{\rm i}\delta_4A_4+rA_1A_2,
\end{eqnarray}
where $A_i$ is the wave amplitude ($i=1,2,3,4$), $A^*_i$ denotes the conjugate
complex, $\nu_1$ is the energy injection coefficient, $\nu_i$ ($i=2,3,4$)
is the dissipation parameter, and $\delta_{3,4}$ are a small mismatch. We
consider $\nu_1=1$, $\nu_2=\nu_3=\nu$, $\delta_4=0$, $\nu_4=-0.83$, transient
time equal to $10^4$, and random initial condition in the interval
$\left[0,1\right]$ for $A_i$, and $A^*_i$. The $r$ parameter controls the
intensity of the fourth wave, at the limit when $r=0$, the four-wave
interaction model becomes the three-wave interaction model.

\begin{figure}[hbt]
  \centering
  \includegraphics[width=0.9\textwidth]{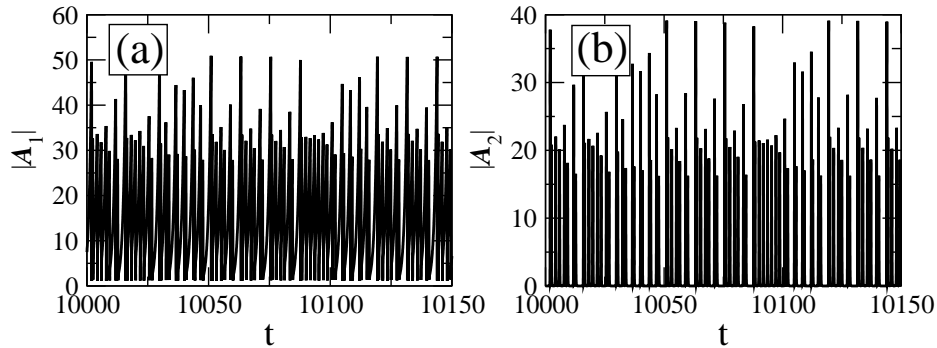}
  \caption{Temporal evolution of (a) $|A_1|$ and (b) $|A_2|$ for $\delta_3=2.2$,
    $\nu=-14.5$, and $r=0$.}
  \label{fig1}
\end{figure}

\begin{figure}[hbt]
  \centering
\includegraphics[width=0.9\textwidth]{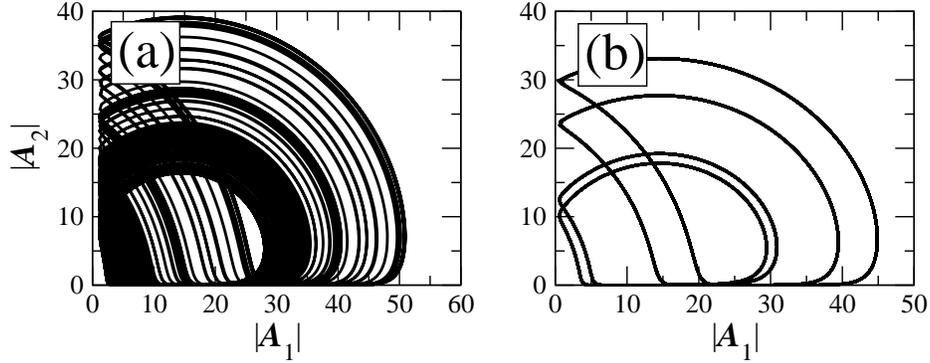}
\caption{(a) Chaotic attractor for $\delta_3=2.2$ and $\nu=-14.5$. (b) Period-4
stable orbits for $\delta_3=1.1$ and $\nu=-14.5$ (periodic evolution). We
consider a transient equal to $10^4$ and $r=0$.}
\label{fig2}
\end{figure}

The temporal evolution of waves $|A_1|$ and $|A_2|$ for the three-wave model
is exhibited in Fig. 1(a) and Fig. 1(b), respectively. The time evolution of
wave $|A_3|$ is similar to $|A_2|$, so it was not shown in Fig. 1. We consider
$\delta_3=2.2$, $\nu=-14.5$, and $r=0$. As a result, the system exhibits
chaotic behaviour, namely the dynamical system is sensitive to initial
conditions. The chaotic attractor is plotted in Fig. \ref{fig2}(a). The
attractor has a large number of periodic orbits identified by varying the
parameter $\delta_3$. Figure \ref{fig2}(b) shows a period-4 stable periodic
orbit for $\delta_3=1.1$ and $\nu=-14.5$. 

\begin{figure}[hbt]
\centering
\includegraphics[width=0.9\textwidth]{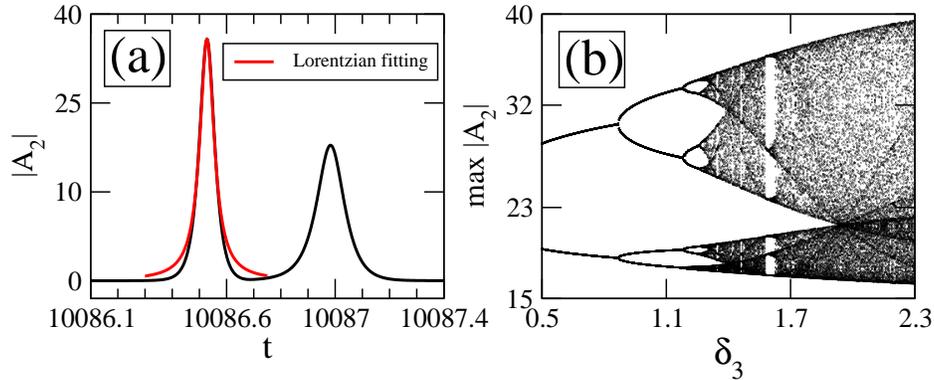}
\caption{(a) Temporal evolution of $|A_2|$ (black line) and its fit by a
Lorentzian function (red line) for $\delta_3=2.2$ and $\nu=-14.5$. (b)
Bifurcation diagram for $\delta_3$.}
\label{fig3}
\end{figure}

The short time interval of the temporal evolution (pul\-se) of $A_2$ (Fig.
\ref{fig1}(b)) can be fitted by a Lorentzian function, as shown in Fig.
\ref{fig3}(a) (red line) for the same values of Fig. \ref{fig1}. Lorentzian
pulses have been associated with chaotic behaviour \cite{maggs12a,maggs12b}. 
In fact, the chaotic behaviour for theses parameters is seen in Fig.
\ref{fig3}(b) where is plotted the bifurcation diagram for the maximum value of
$|A_2|$ as a function of $\delta_3$. The system displays period-doubling,
chaotic behaviour, and periodic windows.

\begin{figure}[hbt]
\centering
\includegraphics[width=0.9\textwidth]{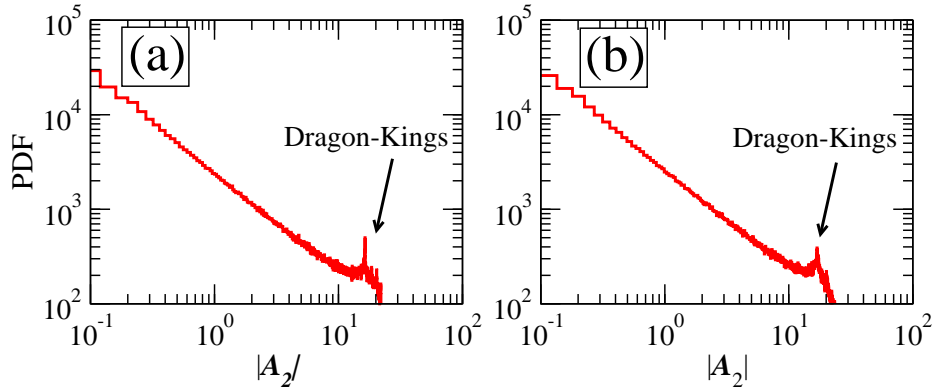}
\caption{Probability density function (PDF) for $r=0$, (a) $\delta_3=2.2$ and
$\nu=-14.5$, and (b) $\delta_3=2$ and $\nu=-15$.}
\label{fig4}
\end{figure}
\section{Dragon-kings death}

Dragon-kings are extreme events that belong to a special class of events. The
extreme events are characterised by power law statistics. However, the
dragon-kings are beyond the power laws. They exhibit humps in the tail of the
distributions for the extended size of events \cite{sornette09}.

We calculate the probability density function (PDF) of the temporal evolution of
$|A_2|$. Figure \ref{fig4} shows the appearance of dragon-kings in the
three-wave interactions model. In Figs. \ref{fig4}(a) and \ref{fig4}(b), we
consider $\delta_3=2.2$ and $\nu=-14.5$, and $\delta_3=2$ and $\nu=-15$,
respectively. The dragon-kings are represented by a lump in the PDF tail.
The events with amplitude above $10$ are outliers and generate the
dragon-kings.

\begin{figure}[hbt]
\centering
\includegraphics[width=0.9\textwidth]{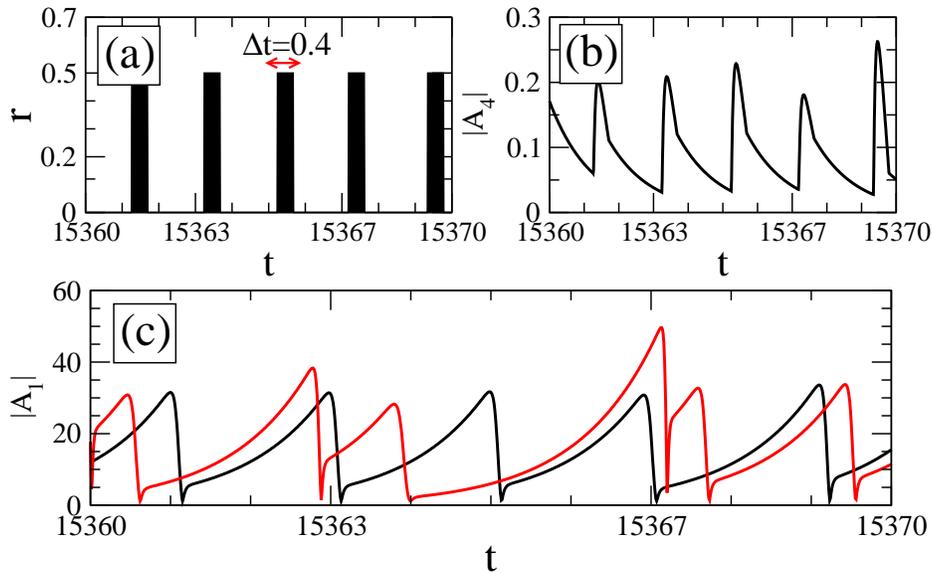}
\caption{(a) $r$ and (b) $|A_4|$ as a function of $t$. (c) Temporal evolution of
$|A_1|$ for $\delta_3=2.2$, $\nu=-14.5$, $r=0$ (red line), $r=0.5$ (black
line), and $\Delta t=0.4$.}
\label{fig5}
\end{figure}

With regard to the four-wave model, the fourth-wave interacts with the first and
second waves. Lopes and Chian \cite{lopes96} used a small sinusoidal wave to
control chaos in the nonlinear three-wave coupling. They demonstrated that a
desired periodic orbit can be obtained applying a fourth wave with small
amplitude. Aiming to suppress the dra\-gon-kings, we apply a fourth wave
($|A_4|$) during a time interval $\Delta t=0.4$ and $r=0.5$ when $d|A_1|/dt>0$,
as shown in Fig. \ref{fig5}(a). Figure \ref{fig5}(b) exhibits the temporal
evolution of $|A_4|$. The energy of the first wave is decreased due to the
application of the fourth wave (Fig. \ref{fig5}(c)). In Fig. \ref{fig5}(c), we
see the behaviour of $|A_1|$ for $\delta_3=2.2$ and $\nu=-14.5$ when $r=0$ (red
line) and $r=0.5$ (black line).

\begin{figure}[hbt]
\centering
\includegraphics[width=0.92\textwidth]{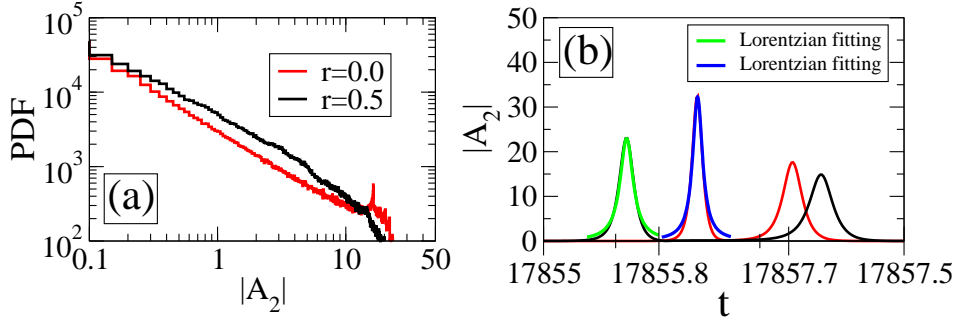}
\caption{(a) Probability density function (PDF) for $\delta_3=2.2$, $\nu=-14.5$,
$r=0$ (red line), $r=0.5$ (black line), and $\Delta t=0.4$. (b) Lorentzian
fitting.}
\label{fig6}
\end{figure}

Figure \ref{fig6}(a) displays the PDF for $r=0$ (red line), $r=0.5$ (black
line), and $\Delta t=0.4$. In our simulations, the fourth wave is able to
produce a small alteration in the PDF. The PDF maintains the same slope, but
the hump in the tail disappears, namely dragon-kings death. In Fig.
\ref{fig6}(b), we see Lorentzian pulses not only in the nonlinear three-wave
interactions (blue line), but also in the wave interactions (green line) where
the dragon-kings are suppressed. Then, a fourth-wave kills the dragon-kings and
does not destroy the Lorentzian pulses.

\section{Conclusions}

In conclusion, we study extreme events in a nonlinear three-wave interactions
model. This model exhibits periodic and chaotic behaviour. In the chaotic
region, we find Lorentzian pulses. The Lorentzian pulses appear in many
physical systems and have been used to characterise chaotic dynamics. Extreme
events are usually rare and the event size distributions follow a power law.
Depending on the system parameters, our simulations show a power law dependence
in the wave amplitude distribution, except for larger values of the amplitude.
The distribution tail has humps, that is an evidence of dragon-kings extreme
events. We consider an external perturbation in the three-wave interactions by
means of a fourth wave. Desired periodic orbits can be achieved applying a wave
with small amplitude. In this work, we show that a fourth wave applied during a
short time can kill dragon-kings. Therefore, the fourth-wave can prevent
catastrophic dragon-kings events in the three-wave interactions. We believe that
in other systems there is a night king, i.e. an external perturbation, that is
able to kill only the dragon-kings. 

\section*{Acknowledgment}
This work was made possible by support from the following Brazilian government
agencies: Funda\c c\~ao Arauc\'a\-ria, CNPq, CAPES, and FAPESP (2011/19296-1,
2015/50122-0, and 2015/07311-7).

\end{document}